%% file: dis_nu.tex
\documentclass[twocolumn,epjc3]{svjour3} 
\usepackage{graphicx}
\usepackage{amsmath}
\usepackage{hyperref}
\usepackage{xspace}
\usepackage[utf8]{inputenc}
\usepackage{color}
\usepackage{todonotes}
\usepackage{lipsum}
\usepackage{cite}
\usepackage[english]{babel}
\usepackage{amssymb}
\usepackage[normalem]{ulem}

\usepackage{subcaption}

\def\beq{\begin{equation}}
\def\beqn{\begin{eqnarray}}
\def\eeq{\end{equation}}
\def\eeqn{\end{eqnarray}}

\newcommand\alphaS{\alpha_\mathrm{s}}

\newcommand\GeV{\ensuremath{\mathrm{GeV}}\xspace}
\newcommand\PeV{\ensuremath{\mathrm{PeV}}\xspace}

\newcommand\mailto[1]{\href{mailto:#1}{#1}}

\newcommand\LOPS{{LO+PS}\xspace}
\newcommand\NLOPS{{NLO+PS}\xspace}

\newcommand\HWpp{{\tt HERWIG++}\xspace}
\newcommand\Herwigpp\HWpp
\newcommand\HERWIGPP\HWpp
\newcommand\PYTHIA{{\tt PYTHIA}\xspace}
\newcommand\PYTHIAn{{\tt PYTHIA~\!\!8}\xspace}

\newcommand\POWHEG{{\tt POWHEG}}
\newcommand\POWHEGBOX{{\tt POWHEG-BOX}\xspace}
\newcommand\POWHEGBOXVTWO{{\tt POWHEG-BOX-V2}\xspace}
\newcommand\RES{{\tt POWHEG-BOX-RES}\xspace}

\newcommand{\xB}{x_{\mathrm{B}}}
\newcommand{\ydis}{y_{\mathrm{DIS}}}

\journalname{Eur. Phys. J. C}
\preprintnumber{CERN-TH-2024-091, MPP-2024-121}

\definecolor{dodgerblue}{rgb}{0.12, 0.56, 1.0}

\definecolor{darkgreen}{rgb}{0.0, 0.5, 0.0}
\definecolor{teal}{rgb}{0.0, 0.5, 0.5}          
\definecolor{purple}{rgb}{0.5, 0.0, 0.5}        

\definecolor{mypink}{rgb}{0.91, 0.33, 0.51}

\begin{document}

\title{An event generator for neutrino-induced Deep Inelastic Scattering 
  and applications to neutrino astronomy}

\author{
Silvia~Ferrario Ravasio\thanksref{addr1,e1}
\and
Rhorry~Gauld\thanksref{addr2,e2}
\and
Barbara J\"ager\thanksref{addr3,e3}
\and
Alexander~Karlberg\thanksref{addr1,e4}
\and
Giulia~Zanderighi\thanksref{addr2,addr4,e5}
}

\thankstext{e1}{\email{\mailto{silvia.ferrario.ravasio@cern.ch}}}
\thankstext{e2}{\email{\mailto{rgauld@mpp.mpg.de}}}
\thankstext{e3}{\email{\mailto{jaeger@itp.uni-tuebingen.de}}}
\thankstext{e4}{\email{\mailto{alexander.karlberg@cern.ch}}}
\thankstext{e5}{\email{\mailto{zanderi@mpp.mpg.de}}}

\institute{%
Theoretical Physics Department, CERN, 1211 Geneva 23, Switzerland\label{addr1}
\and
Max-Planck-Institut f\"ur Physik, Boltzmannstraße 8, 85748 Garching, Germany\label{addr2}
\and
Institute for Theoretical Physics, University of T\"ubingen, Auf der Morgenstelle 14, 72076 T\"ubingen, Germany\label{addr3}
\and
Physik-Department, Technische Universit\"at M\"unchen, James-Franck-Strasse 1, 85748 Garching, Germany\label{addr4}
}

\date{\today}

\maketitle

\abstract{
  We extend the recently presented, fully exclusive,
  next-to-leading-order accurate event generator for the simulation of
  massless neutral- and charged-current deep inelastic scattering
  (DIS) to the case of incoming neutrinos.
The generator can be used to study neutrino-nucleon interactions at
(ultra) high energies, and is relevant for a range of fixed-target
collider experiments and large-volume neutrino detectors,
investigating atmospheric and astrophysical neutrinos.
The matching with multi-purpose event generators such as \PYTHIAn{} is
performed with the \POWHEG{} method, and accounts for parton showering
and non-perturbative effects such as hadronization. 
This makes it possible to investigate higher-order perturbative
corrections to realistic observables, such as the distribution of
charged particles.
To illustrate the capabilities of the code we provide predictions for
several differential distributions in fixed-target collisions for
neutrino energies up to $1~\PeV$.
}
\tableofcontents

\input{sec1-Intro}

\input{sec2-POWHEG}

\input{sec3-Validation}

\input{sec4-Pheno}

\input{sec5-Conclusions}

\section*{Acknowledgments}
We are grateful to Luca Buonocore, Giovanni Limatola, Paolo Nason and
Francesco Tramontano for discussions and to Georg Raffelt for
providing useful references.
In particular, we thank Paolo and Francesco for discussions on the
treatment of collisions involving heavy nuclei.
We also acknowledge stimulating discussions with Andrea Banfi during
early stages of this work.
We are also indebted to Melissa van Beekveld, Eva Groenendijk, Peter
Krack, Juan Rojo, and Valentina Schutze Sanchez for having triggered
this project and tested a pre-release version of our code.
Finally, we are grateful to Alfonso García Soto for advice and
discussions related to experimentally motivated observables.
The work of BJ was supported by the German Research Foundation (DFG)
through the Research Unit FOR 2926.
GZ would like to thank CERN for hospitality while this work was being
finalized.

\appendix
\input{appendix}

\bibliography{dis_nu}

\end{document}

%% file: sec1-Intro.tex
\section{Introduction} \label{sec:intro}

Neutrinos, together with photons, are the most abundant elementary
particles in the universe.
While the properties of photons are extremely well understood, there
are still many outstanding questions regarding neutrinos.
For instance, the origin and nature of neutrino masses (Dirac vs.\ Majorana mass) is not understood, nor is their mass hierarchy (normal vs.\ inverted ordering). 
Furthermore, neutrinos provide a portal to beyond the Standard Model (BSM) physics, making neutrino experiments at the luminosity frontier sensitive to such BSM interactions (see~e.g.~Ref.~\cite{Batell:2009di} for a review).

Neutrino properties are difficult to measure because they only
interact though the weak force.
For this reason, their study often requires large-volume detectors,
which have enabled the discovery of (ultra) high-energy cosmic
neutrinos in 2014, the observation of an astrophysical source of
energetic neutrinos accompanied by gamma-ray emissions in 2018, and
the determination of the oscillation properties of multi-$\GeV$ energy
atmospheric neutrinos (see e.g. Ref.~\cite{Ackermann:2022rqc} for a
review of these and several recent results).
Ongoing experiments such as ANTARES~\cite{ANTARES:2011hfw},
Baikal~\cite{BAIKAL:2005qnn}, IceCube~\cite{IceCube:2016zyt}, and
KM3NeT~\cite{KM3Net:2016zxf}, will continue to extract information on
(ultra) high-energy neutrinos to which their detectors are exposed.
Moreover, a range of proposed next-ge\-ne\-ra\-tion detectors will facilitate precise measurements of (ultra) high-energy neutrinos from atmospheric and cosmic sources. This advancement will usher in a new era of precision, enabling to probe neutrino properties, their interactions, and fundamental symmetries at the highest possible energies.
Furthermore, this
programme will be instrumental to discover and characterize the
astrophysical sources of the most energetic cosmic and gamma-rays.

Additional data opportunities come from high-lumi\-nosity experiments.
For example, measurements of neu\-trino-matter scattering at collider
facilities (e.g. charm production measured by
NuTeV~\cite{NuTeV:2007uwm}) have provided important information on the
hadron structure.
Forward-physics facilities such as
SND@LHC~\cite{SNDLHC:2022ihg,SNDLHC:2023pun},
SHiP~\cite{SHiP:2015vad}, and
FASER$\nu$~\cite{FASER:2019dxq,FASER:2020gpr,FASER:2023zcr}, are
already taking data and the Forward Physics Facility (FPF) is on the
horizon for the HL-LHC~\cite{Anchordoqui:2021ghd,Feng:2022inv}.
A major goal of each of these experiments is to extract the flavour
and energy-dependence of the neutrino flux to which their detector is
exposed. This requires, in addition to a detailed understanding of the
detector, precise knowledge of the expected differential rates of
neutrino-nu\-cleon scattering for varying neutrino flavour and energy.

At the large energies under consideration (multi-$\GeV$ and above),
the scattering rate of neutrinos with matter is dominated by the deep
inelastic scattering (DIS) process.
The role of theory in this context is thus an important one: it
provides a well defined (and rigorously tested) computational
framework, that of collinear factorisation~\cite{Collins:1989gx}, to
predict the differential scattering rates of neutrinos.
This framework is reliable provided the exchanged momentum, $Q^\mu$,
satisfies $|Q^2| \gtrsim m_p^2$, $m_p$ being the proton mass, and can
be applied across many orders of magnitude in neutrino energy. It
relies on a combination of perturbative QCD ingredients, and of the
knowledge of the universal partonic content of the colliding hadrons
(as extracted from global analyses of hadron collider data), see
Ref.~\cite{Ethier:2020way} for a recent review.

This theoretical framework can be straightforwardly applied to the
case of (ultra) high-energy neutrino-nu\-cleon scattering by
expressing the differential cross-sec\-tion in terms of DIS structure
functions (see for example the discussion in Section II
of~\cite{Cooper-Sarkar:2011jtt}).
The structure functions encapsulate the strong dynamics of the nucleon
as struck by an exchanged gauge boson, and they can be predicted
through the convolution of parton distribution functions (PDFs) with a
set of perturbatively calculated coefficient functions. The simplicity
of this approach stems from the fact that the structure functions
provide an inclusive description of all QCD radiation in the
scattering process.
On the other hand, it is limited as predicted cross-sections are
differential only in quantities inclusive over QCD radiation, such as
the leptonic momentum transfer $Q^2$ and the Bjorken momentum
fraction, $\xB$.
The massless hard coefficient functions that enter into the structure
functions have been computed at
3-loops~\cite{SanchezGuillen:1990iq,vanNeerven:1991nn,Zijlstra:1991qc,Zijlstra:1992qd,Zijlstra:1992kj,vanNeerven:1999ca,vanNeerven:2000uj,Moch:1999eb,Moch:2004xu,Vermaseren:2005qc,Vogt:2006bt,Moch:2007rq,Davies:2016ruz,Blumlein:2022gpp}.
Following the structure-function approach, dedicated theoretical
studies of neutrino-nucleon DIS at high energies have appeared over
the years, both at leading-order
(LO)~\cite{Gandhi:1998ri,Gluck:1998js,Cooper-Sarkar:2007zsa,Connolly:2011vc},
next-to-leading order (NLO)~\cite{Cooper-Sarkar:2011jtt} and recently
at next-to-next-to-leading order (NNLO) in
QCD~\cite{Bertone:2018dse,Xie:2023suk}.
The impact of the physics effects due to heavy-quark masses, nuclear
modifications of PDFs, and resummation of small-$x$ contributions has
been studied in Refs.~\cite{Bertone:2018dse,Xie:2023suk}, the role of
certain classes of QED effects has been investigated in
Refs.~\cite{Seckel:1997kk,Alikhanov:2015kla,Gauld:2019pgt,Zhou:2019vxt,Zhou:2019frk,Xie:2023qbn},
and effects beyond collinear factorisation have also been
discussed~\cite{Jalilian-Marian:2003ghc,Fiore:2005wf,Block:2013nia,Albacete:2015zra,Goncalves:2015fua,Arguelles:2015wba}.

Predictions obtained in this way provide an important benchmark for
differential DIS cross-sections in terms of QCD-inclusive quantities
(e.g.\ distributions of $Q^2$ and $\xB$), as well as the total
cross-section. However, they do not provide an exclusive description
of the radiation which is generated in the scattering process.
This is a significant limitation for many analyses at current (and
future) neutrino experiments which aim to reconstruct the energy and
direction of the incoming neutrino, and which rely on an accurate
description of the properties of final-state radiation (such as the
distribution of electromagnetically charged and neutral particles) to
do so.
A step towards overcoming this issue is made in the current work with
the development of an event generator for the simulation of
neutrino-induced massless neutral- and charged-current DIS based on
the \POWHEG~\cite{Nason:2004rx,Frixione:2007vw} method. The predictions
obtained with this program are accurate at NLO in QCD and can be
matched with a multi-purpose Shower Monte Carlo generator to provide a
fully exclusive description of the scattering process. The
implementation is based on the existing generator for charged-lepton
induced DIS processes presented in~\cite{Banfi:2023mhz}, and has been
implemented in the publicly available framework
\RES~\cite{Jezo:2015aia}. The code can be obtained from
\url{svn://powhegbox.mib.infn.it/trunk/User-Processes-RES/DIS}.

While this paper was being finalised, an NLO accurate event generator
implementation for lepton-hadron DIS was
presented~\cite{Buonocore:2024pdv}. This implementation is based on
the \POWHEGBOXVTWO{} framework~\cite{Alioli:2010xd}, and has a
particular focus on processes with a heavy lepton, such as a tau
neutrino, and/or a heavy charm quark in the final state. We briefly
discuss the differences between the two codes in~\ref{app:compothers}.

The structure of the paper is as follows: in Sec.~\ref{sec:POWHEG} we
summarise the main details of the process implementation and new
features as compared to the existing generator which describes
charged-lepton induced DIS; a validation of the code for various DIS
subprocesses is provided in Sec.~\ref{sec:Validation}; in
Sec.~\ref{sec:Pheno} we present phenomenological results for
several distributions of charged particles and charmed hadrons for
incident neutrino energies of $10^5$ and $10^6~\GeV$. Concluding
remarks are presented in Sec.~\ref{sec:Conclusions}.
A complete list of all the new features in the code, and how to
use them, is provided in~\ref{sec:processes}.

%% file: sec2-POWHEG.tex
\section{Details of the implementation}
\label{sec:POWHEG}
In this section we discuss the extensions needed to augment the \RES
generator for massless neutral- and charged-current DIS, presented in
Ref.~\cite{Banfi:2023mhz}, to allow for the inclusion of initial-state
neutrinos and generic (massive) nuclear targets. The \RES framework
combines NLO-QCD calculations with parton showers (PS) according to
the \POWHEG{} method, and was originally only designed to handle
hadron-hadron collisions. One of the main novelties of
Ref.~\cite{Banfi:2023mhz} was the design of new momentum mappings that
preserve the special kinematics of DIS in the FKS subtraction
formalism~\cite{Frixione:1995ms,Frixione:2007vw} as implemented in the
\RES framework.

The original generator of Ref.~\cite{Banfi:2023mhz} was designed to
describe DIS reactions resulting from the collision of a massless
proton with a charged lepton, relevant to interpret data from, for
instance, HERA and the forthcoming Electron Ion Collider (EIC). It was
since extended to also include polarised beams in
Ref.~\cite{Borsa:2024rmh}.

The extension presented here contains three new major features: 1. The
incoming lepton can now be of any species, in particular it can be a
neutrino or a charged lepton; 2. The code can now handle a massive
nucleon at rest, of relevance to fixed-target experiments; 3. A
variable flux can be supplied for the incoming lepton beam. The
handling of massive nucleons at rest is described in
Sec.~\ref{sec:fixed-target}, and a discussion of how to consistently
account for the nuclear target PDFs can be found in
Sec.~\ref{sec:target}. Although in this paper we focus on
phenomenological studies of neutrino beams with fixed energy, we
discuss how to include a variable flux in
Sec.~\ref{sec:neutrino-flux}.
Finally in Sec.~\ref{app:compothers} we comment on our momentum
mappings and how mass effects are approximately included.

\subsection{Fixed-target experiments}
\label{sec:fixed-target}
By default, the \RES{} can only handle collisions of massless
beams. In this section we therefore describe how to perform
fixed-target collisions, using a set of massless beams.
Denoting the energies of two massless colliding beams in the laboratory frame
by \(E_1\) and \(E_2\), the \POWHEGBOX{} builds the four-momenta of
the beam particles as follows:
\begin{align}
   \label{eq:powheg-beams}
  k_{\rm beam, 1} =& \left\{ E_1, 0, 0, +E_1 \right\}, \notag\\
  k_{\rm beam, 2} =& \left\{ E_2, 0, 0, -E_2 \right\}. 
\end{align}
These four-vectors are then used to construct the momenta of the
incoming elementary fermions entering the scattering process.

To account for the collision of a beam of massless particles of energy
\( E \) with a fixed target nucleon (i.e. proton or neutron) of mass
\(m \) we extend this approach by effectively treating the nucleon as
massless. 
In the fixed-target frame the true momenta are given by the lepton
beam momentum, $P_1$, and the fixed target momentum, $P_2$,
\begin{align}
 P_1 =& \left\{ E , 0, 0,  E  \right\},\notag \\
 P_2 =& \left\{ m, 0, 0, 0\right\}.
\end{align}
From these momenta we obtain a centre-of-mass energy, $E_\mathrm{CM}$, via 
\begin{equation}
  E_\mathrm{CM}^2 = (P_1+P_2)^2 = 2mE +m^2.
\end{equation}
We then trivially observe that if we pick $E_1=E_2=E_{\mathrm{CM}}/2$
in Eq.~\eqref{eq:powheg-beams} we can construct a set of massless
momenta that coincide with the centre-of-mass frame of the
fixed-target collision. Now consider the boost from the centre-of-mass
frame to the \emph{true} fixed-target frame. Applying this boost to
our newly constructed massless momenta we can construct massless beam
momenta in Eq.~\eqref{eq:powheg-beams} where the energies of the beams
are set to
\begin{align}
  E_1 = E + m/2, \qquad E_2 = m/2.
\end{align}
Both the massless centre-of-mass and massless fixed-target momenta satisfy \(k_{\rm beam, 1} + k_{\rm beam, 2} =
P_1+P_2\) by construction, but do not preserve the mass of $P_2$. In practice we expect the massles construction to be reliable as long as $m/E \ll 1$.
The two sets of momenta result in equivalent predictions, since they are
related by a boost, but in practice we find that using the
centre-of-mass momenta is numerically more stable for ultra-high
energy collisions ($E/m \gtrsim 10^5 - 10^6$). We provide both options
in the code, as described in \ref{sec:processes}.

We note that when interfacing the events to the parton shower,
e.g. \PYTHIAn{}, the actual mass of the nucleon is restored while
retaining the centre-of-mass energy of the two beams, thereby
restoring the correct kinematics.

\subsection{Nucleon targets}
\label{sec:target}
When considering lepton scattering off the nucleons of a bound nucleus,
it is important to differentiate whether the nucleon target is a
proton or a neutron. This distinction is relevant for the eventual
matching to the parton shower, where the quantum numbers of the
nucleon remnant must be known.
The selection of the nucleon type in the \texttt{powheg.input} file
can be made by setting the integer \verb|ih2|, as described
in~\ref{sec:processes}.
For the selection of a neutron, we provide the option to either
directly use neutron PDFs, or to instead provide a set of proton PDFs
which the program then internally converts via an isospin
transformation.
The latter option has been added because some nuclear PDF fitting
groups (which assume isospin symmetry) provide the nuclear PDFs in the
format of average bound proton PDFs.

Taking as an example the scattering of neutrinos with H$_2$O molecules,
the total cross section is given by
\begin{align}
  \label{eq:h2o-xsec}
\sigma^{\text{H}_2\text{O}}_{\nu} = 2 \sigma^{p}_{\nu} + Z \sigma^{p/O}_{\nu} + (A-Z) \sigma^{n/O}_{\nu}\,,
\end{align}
where $\sigma^{p}_{\nu}$, $\sigma^{p/O}_{\nu}$, and
$\sigma^{n/O}_{\nu}$ are the cross sections for free protons, bound
protons and bound neutrons, respectively, and $Z=A-Z=8$ for oxygen. In
this case one has to perform three different runs: The first run using
free protons, the second using bound protons, and the third using
bound neutrons. For both the bound protons and neutrons one should use
nuclear PDFs. The final showered result is then given by combining
these three runs according to the above equation.

When considering scattering on a single nucleus (such as oxygen), 
one could generate events using a PDF which is the 
appropriate admixture of protons and neutrons in the target nucleus.
This would then require two instances of the parton shower -- one for
the proton and one for the neutron -- that one selects event by event
with the probability determined by the relative fraction of the PDFs
for protons and neutrons in the nucleus.
For an extension of the \PYTHIAn{} Monte Carlo event generator that
enables the simulation of collisions between a generic hadron beam on
a generic nuclear target see Ref.~\cite{Helenius:2024vdj}. That work
combines the extension of \PYTHIAn{} to deal with heavy ion
collisions~\cite{Bierlich:2018xfw}, and the extension to collisions of
a varying hadron beam on a proton target~\cite{Sjostrand:2021dal}.

\subsection{Variable neutrino flux}
\label{sec:neutrino-flux}
By default, we consider a monochromatic incoming lepton flux.
To account for the typical environment of a neutrino-induced DIS
process our new implementation additionally provides an option for a
realistic neutrino flux.
The user can implement a realistic flux by modifying the function
\texttt{pdf\_lepton\_beam}, which is contained in the file
\texttt{lepton\_flux.f}.
If importance sampling associated with the lepton's energy fraction is
required, the user can modify the function \texttt{sample\_x\_lepton},
also contained in the same file. This function builds the lepton's
energy fraction given a random number.

The correct modeling of such a flux depends on the specific experiment
and goes beyond the scope of this publication.
A detailed study for SND@LHC, FASER$\nu$, and the planned FPF
experiments FLArE and FASER$\nu$2, using our code and framework, will
be presented in Ref.~\cite{RojoDIS}.

\subsection{On the momentum mappings, mass effects and possible extensions to more complex processes }
\label{app:compothers}
  In Ref.~\cite{Banfi:2023mhz} we introduced new momentum mappings,
  focusing on the fully massless case, and used them to implement a
  DIS generator in the \RES{} fra\-me\-work.
  A \POWHEGBOXVTWO{} generator was presented in
  Ref.~\cite{Buonocore:2024pdv}, where such mappings have been
  generalised to account for an explicit lepton-mass dependence.
This mass dependence can be relevant when studying processes involving
$\tau$ leptons for $Q$ values not much higher than the mass of the
$\tau$ lepton, as probed by the FASER$\nu$ and SHiP experiments.
Additionally, the initial-state map of Ref.~\cite{Buonocore:2024pdv}
supports heavy coloured final-state particles.
In Ref.~\cite{Buonocore:2024pdv} there is no dedicated treatment of
the collinear singularities associated with the emissions from a
final-state heavy quark. This would have required an extension of the
work of Refs.~\cite{Barze:2012tt,Buonocore:2017lry} to the DIS case.
Instead, contributions associated with emissions collinear to a heavy
quark, as well as power-suppressed terms, are included at fixed-order
accuracy as a separate regular contribution, involving potentially
large mass logarithms.
Therefore, when the centre-of-mass energy becomes very large
relative to the relevant quark masses - as is the case in
(ultra) high-energy neutrino collisions -- the massless QCD
calculation, available in both codes, has to be preferred.
Indeed we stress that, even in the massless approximation, 
when generating radiation in \POWHEG{}, mass thresholds for the
heavy-quarks are present so that the leading mass-logarithms
associated with collinear final-state emissions are included to all
orders.
Therefore, in \POWHEG{} events, radiation with a transverse momentum
smaller than the mass of the emitting quark is vetoed, effectively
mimicking a dead cone.
Furthermore, we also stress that even for calculations where
final-state quarks or leptons are treated as massless in the
matrix-elements, the generated momenta of the \POWHEG{} events are
reshuffled to include finite masses and that the subsequent parton
shower is fully aware of mass effects, including the correct decays of
$\tau$ leptons.

We also note that, in the massless limit, the maps of
Refs.~\cite{Banfi:2023mhz,Buonocore:2024pdv} as well as
the handling of final-state radiation are identical.
For initial-state radiation instead, while the kinematic map is
the same, they differ in the definition of the \POWHEG{} hardness
parameter away from the soft and collinear limits.
Denoting by $\xi$ and $y$ the energy-fraction and the cosine of the
emission angle and by $\bar s$ the centre-of-mass energy of the
underlying Born, the two definitions are given by
\begin{align}
 \quad t_{\rm ISR} &= \frac{\xi^2}{2-\xi(1+y)} \bar s (1-y), &\text{in Ref.\cite{Banfi:2023mhz}},\\
  \quad t_{\rm ISR} &= \frac{\xi^2}{2(1-\xi y)} \bar s (1-y), &\text{in Ref.\cite{Buonocore:2024pdv}}.
\end{align}
It is evident that the two definitions are identical in the soft ($\xi \to
0$) and in the collinear ($y \to 1$) limits. We thus conclude that the
two codes have the same formal accuracy.

The \RES{} framework, specifically design\-ed to handle hadronic
scattering processes that contain decaying resonances and thus require
the inclusion of radiative corrections not only in the production, but
also in the decay process, is particularly well-suited for extending
our approach to other processes relevant for the phenomenology of
hadron-hadron collisions, as well as including electroweak corrections
in DIS.
In particular, for processes such as vector boson fusion or vector
boson scattering, that can be modelled as generalised two-fold DIS
processes, the \RES{} framework is best suited to handle the two
hadronic sub-sectors with a factorised approach.
In this sense, our \RES implementation of the genuine DIS process
provides a stepping stone towards the development of suitable
generators for such more complex hadron-hadron collision processes.
It is also more straightforward to include soft
photon emissions connecting the leptonic and the hadronic sectors of
the DIS process in the \RES framework.
This feature will be essential for the inclusion of electroweak
corrections in the generator.

%% file: sec3-Validation.tex
\section{Fixed-order validation}
\label{sec:Validation}
To validate our new implementation, we perform a comparison with
existing fixed-order predictions for selected DIS processes where a
neutrino is scattering off an oxygen target.
Specifically, we compute the quantity
\begin{align}
  \label{eq:o-xsec}
\sigma^{\text{i/O}}_{\nu} = Z/A \,\sigma^{p/O}_{\nu} + (A-Z)/A \, \sigma^{n/O}_{\nu}\,,
\end{align}
which is the per-nucleon cross-section for an (isoscalar) oxygen target.

In our work, we have used the set of nuclear
PDFs \\\verb|nNNPDF30_nlo_as_0118_p_O16|~\cite{AbdulKhalek:2022fyi},
which is provided in a variable flavour number scheme ($n_f^{\rm max}
= 5$) and is expressed in terms of average bound-proton states.
We note that top quark contributions to DIS are expected to be
negligible below neutrino energies of about $1$ PeV. If higher
energies are considered, the inclusion of the top-quark contributions
could become relevant for the CC process, see
Ref.~\cite{Garcia:2020jwr}.
We generated separately a sample for proton ($p$) and neutron ($n$)
targets as described in Sec.~\ref{sec:target}.
The neutron PDF is obtained from the proton one using isospin
relations, as described in \ref{sec:processes}.
The central renormalisation $\mu_R$ and factorisation scales $\mu_F$ are set to the momentum transfer $Q$.
Scale uncertainties are estimated by performing an independent
variation of $\mu_R$ and $\mu_F$ by a factor $2$ up and down, subject
to the constraint $1/2 \leq \mu_R/\mu_F \leq 2$.
We impose a lower cutoff on $Q$ of $Q_{\rm min} = 2.0~\GeV$ which
ensures the PDFs and the strong coupling $\alphaS$ are never evaluated
at scales below $1.0~\GeV$.

For the masses and widths of the electroweak gauge bosons we start
from the on-shell values given in the
PDG~\cite{ParticleDataGroup:2022pth}
\begin{align}
&m_W^{\rm OS} = 80.3770~\GeV\,,\qquad
\Gamma_W^{\rm OS} = 2.085~\GeV\,,\nonumber\\
&m_Z^{\rm OS} = 91.1876~\GeV\,,\qquad
\Gamma_Z^{\rm OS} = 2.4955~\GeV\,, 
\end{align}
and convert them to the pole values as described e.g.\ in
Ref.~\cite{Denner:2019vbn}, which are then used as input values for
the simulations.  For the Fermi constant and the weak mixing angle we
use
\beq
G_{F} \!= \!1.1663787\!\times\! 10^{-5} ~\GeV^{-2}\!,\;\;
\sin^2 \theta_W = 0.2316.
\eeq
The value of electromagnetic coupling $\alpha$ is derived from these parameters as
$\alpha = \sqrt{2}/\pi G_F m_W^2  \sin^2 \theta_W$.
For the charged current process this effectively implies the
replacement $\alpha/\sin^2 \theta_W\to G_{F}^2 m_W^2$ when evaluating
the squared amplitude. This choice ensures the resummation of the
leading universal electroweak corrections~\cite{Denner:1991kt}.
A similar replacement also takes place for the neutral current
process, while additional dependencies on $\sin^2 \theta_W$ appearing
in the squared amplitude are described by our chosen value of $\sin^2
\theta_W$ (which is fixed to the measured effective weak mixing
angle).
This approach provides an accurate normalisation of the couplings, and
ensures that the measured on-shell values of the boson masses enter
the propagators for both the charged and neutral current processes we
are describing.

For the entries of the Cabibbo-Kobayashi-Maskawa matrix we have used
\begin{align}
  &V_{ud} = V_{cs} = 0.97446\,, \nonumber \\
  &V_{us} = V_{cd} = 0.22456\,, \nonumber \\
  &V_{tb} =1\,,
\end{align}
with all other entries zero.

The fixed-order predictions are provided at both NLO and NNLO, and
have been obtained using the implementation
from~\cite{Bertone:2018dse}, which relies on {\tt
  APFEL}~\cite{Bertone:2013vaa} for the computation of the DIS
structure functions up to
NNLO~\cite{vanNeerven:1991nn,Zijlstra:1991qc,Zijlstra:1992qd,Zijlstra:1992kj,Moch:1999eb}. In each case the same NLO accurate nuclear PDF set specified  above is used.
The structure functions have been benchmarked against {\sc
  Hoppet}~\cite{Salam:2008qg,Bertone:2024dpm} and the fixed-order
predictions have been cross-checked against predictions from {\tt
  disorder}~\cite{Karlberg:2024hnl}.

In the following we denote by \LOPS and \NLOPS predictions at LO and
NLO, respectively, matched to parton shower.  For the \NLOPS
predictions shown below we interface our \POWHEGBOX{} implementation
to \PYTHIA 8.308~\cite{Bierlich:2022pfr}, with default settings
(Monash tune~\cite{Skands:2014pea}), and we use the simple shower with
fully-local recoil option~\cite{Cabouat:2017rzi}.
For the results presented in this section, QED radiation and
hadronization effects are not included.

\begin{figure*}[t]
  \centering
  \begin{subfigure}[h]{.40\textwidth}
    \includegraphics[width=\textwidth]{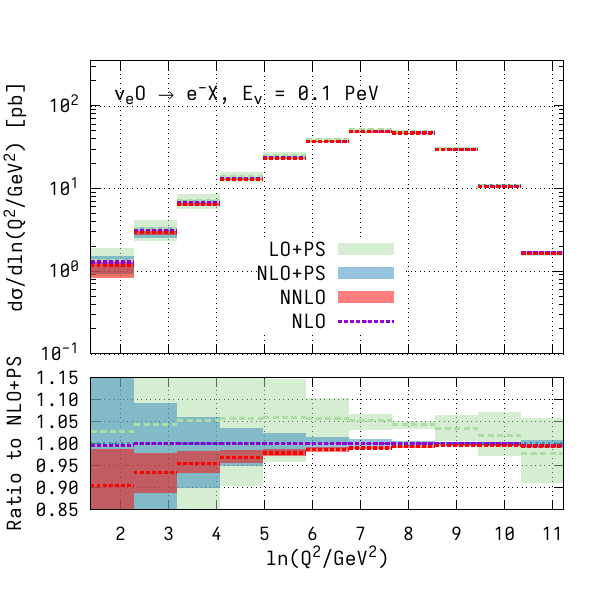}
    \subcaption{
      }
    \label{fig:CCQ2}
  \end{subfigure}
  \hspace{1cm}
  \begin{subfigure}[h]{.40\textwidth}
     \includegraphics[width=\textwidth]{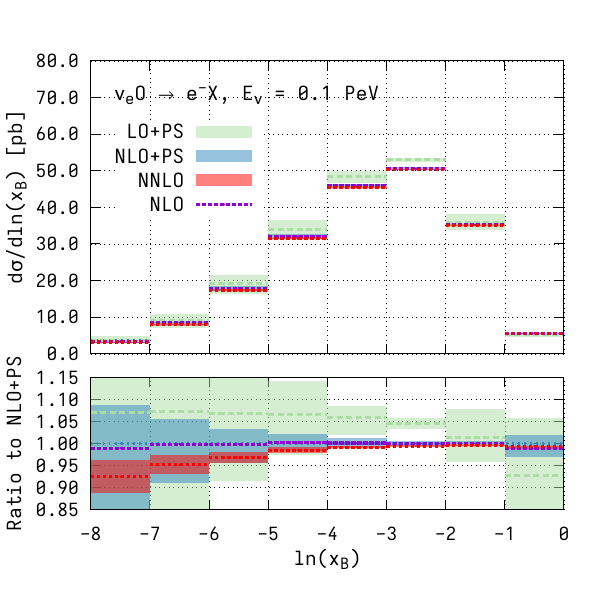}
    \subcaption{}
    \label{ffig:CCXbjk}    
  \end{subfigure}
  \caption{Differential cross-section (per-nucleon) for the charged-current
    scattering of a neutrino $\nu_e$ of energy $E_{\nu} = 0.1~\PeV$ on
    oxygen, with respect to $\ln(Q^2/{\rm GeV}^2)$ (left) and
    $\ln(\xB)$ (right) at LO+PS (green), NLO+PS (blue), pure NLO
    (violet) and NNLO (red).
    The widths of the bands indicate scale uncertainties estimated by
    a 7-point variation of $\mu_R$ and $\mu_F$ by a factor of two
    around the central value $Q$. The lower panels show ratios to the
    respective \NLOPS results with $\mu_R=\mu_F=Q$.  }
  \label{fig:validation_CC}
\end{figure*}

\begin{figure*}[t]
  \centering
  \begin{subfigure}[h]{.40\textwidth}
    \includegraphics[width=\textwidth]{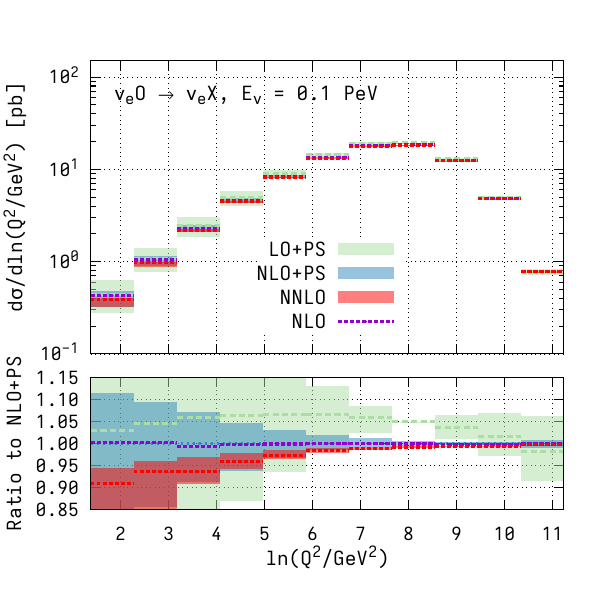}
    \subcaption{}
    \label{fig:NCQ2}
  \end{subfigure}
  \hspace{1cm}
  \begin{subfigure}[h]{.40\textwidth}
     \includegraphics[width=\textwidth]{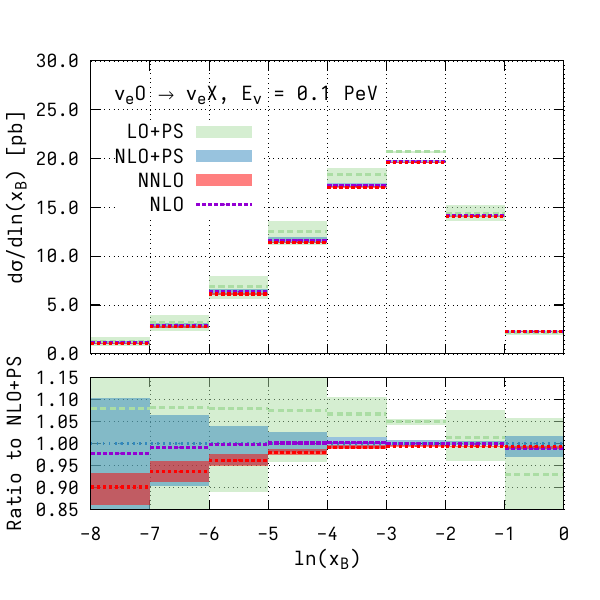}
    \subcaption{}
    \label{fig:NCXbjk}    
  \end{subfigure}
  \caption{Analogous to Fig.~\ref{fig:validation_CC} for the neutral current  process $\nu_e O \to \nu_e X$.}
  \label{fig:validation_NC}
\end{figure*}

We have performed comparisons of cross sections differential with
respect to the DIS variables $Q^2$ and $\xB$ with different neutrino
energies for both charged current (CC) and neutral current (NC)
processes in the case of either incoming neutrinos or antineutrinos
for the scattering off an oxygen target at rest, i.e.\ the reactions
$\nu_e O \to e^- X$, $\bar \nu_e O \to e^+ X$, $\nu_e O \to \nu_e X$,
and $\bar \nu_e O \to \bar \nu_e X$, where $X$ denotes the unresolved
hadronic final state of the DIS reaction.
We show explicit results for the selected processes $\nu_e O \to e^-
X$ and $\nu_e O \to \nu_e X$ in Fig.~\ref{fig:validation_CC} and
Fig.~\ref{fig:validation_NC}, respectively.  In both cases we consider
fixed-target collisions with a neutrino energy of $E_{\nu} =
0.1~\PeV$, corresponding to a neutrino-nucleon centre-of-mass energy
of $\sqrt{s} = 431.74~\GeV$. In Figs.~\ref{fig:validation_CC}
and~\ref{fig:validation_NC} we show the differential results with
respect to $\ln(Q^2/{\rm GeV}^2)$ (left panel) and $\ln(\xB)$ (right
panel) for CC and NC, respectively.
For the \LOPS, \NLOPS and NNLO predictions, we show scale variation
uncertainties, while statistical errors are much smaller and not shown
here.

\begin{table*}[h]
  \centering
\renewcommand{\arraystretch}{1.5}  
  \scalebox{1.2}{
    \begin{tabular} { | l | r | r | }
    \hline
    \multicolumn{3} { | c | }{Cross-sections with the cut $Q>2~\GeV$ for $E_\nu=0.1$ PeV}\\
    \hline
    \multicolumn{1} { | c | }{Process}    & \multicolumn{1} { | c | }{\NLOPS (pb)} & \multicolumn{1} { | c | }{NNLO (pb)} \\
    \hline
$\nu_e O \to e^- X$ 
& $200.68^{+2.87}_{-3.53}\,\text{(scales)}\,^{+2.68}_{-3.29}\,\text{(PDFs)}$ & $197.92^{+1.21}_{-1.02}\,\text{(scales)}$ \\
$\bar \nu_e O \to e^+ X$ 
& $168.32^{+2.73}_{-3.34}\,\text{(scales)}\,^{+2.64}_{-3.34}\,\text{(PDFs)}$ & $ 165.73^{+1.16}_{-0.99}\,\text{(scales)}$ \\ \hline
 $\nu_e O \to \nu_e X$ 
 & $75.97^{+1.25}_{-1.39}\,\text{(scales)}\,^{+0.76}_{-0.91}\,\text{(PDFs)}$ & $ 74.81^{+0.44}_{-0.41}\,\text{(scales)}$ \\
$\bar \nu_e O \to \bar \nu_e X$ 
& $64.85^{+1.21}_{-1.33}\,\text{(scales)}\,^{+0.78}_{-0.82}\,\text{(PDFs)}$ & $ 63.75^{+0.42}_{-0.40}\,\text{(scales)}$ \\
    \hline   
    \end{tabular}
    }
    \caption{Total cross-section with the cut $Q>2~\GeV$ for a
      selection of DIS processes with a (anti-)neutrino of energy
      $E_\nu=0.1$ PeV at \NLOPS and NNLO accuracy. The quoted
    uncertainties are due to scale variation. For the \NLOPS results
    we also indicate the size of the PDF uncertainties in the second
    entry.}
    \label{tab:sigmaE5}
\end{table*}

\begin{table*}[h]
  \centering
\renewcommand{\arraystretch}{1.5}  
  \scalebox{1.2}{
    \begin{tabular} { | l | r | r | }
    \hline
    \multicolumn{3} { | c | }{Cross-sections with the cut $Q>2~\GeV$ for $E_\nu=1$ PeV}\\
    \hline
    \multicolumn{1} { | c | }{Process}    & \multicolumn{1} { | c | }{\NLOPS (pb)} & \multicolumn{1} { | c | }{NNLO (pb)} \\
    \hline
$\nu_e O \to e^- X$ & $624.49^{+14.14}_{-16.44}\,\text{(scales)}\,^{+15.26}_{-15.42}\,\text{(PDFs)}$ & $ 613.42^{+5.02}_{-3.70}\,\text{(scales)}$ \\
$\bar \nu_e O \to e^+ X$ & $598.05^{+14.00}_{-16.33}\,\text{(scales)}\,^{+15.81}_{-15.90}\,\text{(PDFs)}$ & $ 587.09^{+4.99}_{-3.68}\,\text{(scales)}$ \\ \hline
$\nu_e O \to \nu_e X$ & $258.59^{+6.48}_{-7.11}\,\text{(scales)}\,^{+5.67}_{-5.69}\,\text{(PDFs)}$ & $ 253.61^{+2.06}_{-1.61}\,\text{(scales)}$ \\
$\bar \nu_e O \to \bar \nu_e X$ & $248.73^{+6.43}_{-7.07}\,\text{(scales)}\,^{+5.82}_{-5.58}\,\text{(PDFs)}$ & $ 243.78^{+2.05}_{-1.60}\,\text{(scales)}$ \\
    \hline   
    \end{tabular}
    }
    \caption{Analogous to Tab.~\ref{tab:sigmaE5}, now for $E_\nu=1$ PeV.}
    \label{tab:sigmaE6}
\end{table*}
    
We observe that at low-to-moderate values of $Q^2$, within the given
scale uncertainties, the fixed-order NLO predictions agree with the
\NLOPS results and are very similar to the \LOPS results.  Obviously,
the impact of higher-order corrections is small on this observable.
For the Bjorken variable we find agreement between the NLO and the
\NLOPS results, as expected for this inclusive quantity.
Technically we expect the agreement between NLO and \NLOPS to be
near-perfect, as the shower without QED radiation preserves the lepton
momenta.
However, as discussed in Sec.~\ref{app:compothers}, the \POWHEGBOX{}
performs a small momentum reshuffling to account for the finite quark
and lepton masses, and additionally, as was discussed in
Sec.~\ref{sec:fixed-target}, at event level the nucleon mass is
restored.
This reshuffling has a tiny impact on the $Q^2$ and $\xB$
distributions, as was also discussed in Ref.~\cite{Banfi:2023mhz}.

It is worth noticing that the \NLOPS result is not always contained
with in the scale variation band of the \LOPS result.
The perturbative uncertainties of the \LOPS result are not expected to
be fully covered by a standard scale variation, as at this order only
$\mu_F$ can be varied, while $\mu_R$ does not even enter.
On the other hand, we see that the NNLO prediction is fully contained
within the scale variation band of the \NLOPS prediction, thereby
establishing confidence in the reliability of our prediction.

In addition to the differential validation, we also report results for
the per-nucleon cross section, with a cut $Q\ge 2$ GeV, obtained up to NNLO
accuracy in Tab.~\ref{tab:sigmaE5} for $E_\nu=0.1$ PeV and
Tab.~\ref{tab:sigmaE6} for $E_\nu=1$ PeV.
The results are given for a selection of processes and (anti)-neutrino
energies.
The central prediction and the uncertainty due to scale variations are
shown in each case.  It has been checked that the NLO entries obtained
with this generator (labelled as \NLOPS) reproduce exactly, including
scale variations, the NLO results based on the structure function
computation.  For that reason we only show the \NLOPS results.
We have additionally reported the uncertainties due to the nuclear
PDFs computed at NLO.
Typically these uncertainties are in the range of $(1-2)\%$ and are
similar in size to those of the scale uncertainties at NLO.
Finally, we note that the structure functions are non-zero below $Q_{\rm min}$
(and hence so is the cross-section), but the description of this
region goes beyond the applicability of collinear
factorisation. Alternative (data-driven) approaches exist to describe
the low-$Q$ region, see for
example~\cite{Bodek:2002vp,Bodek:2003wd,Bodek:2004pc,Bodek:2010km,Bodek:2021bde}
and, more recently, Ref.~\cite{Candido:2023utz}.

%% file: sec4-Pheno.tex
\section{Phenomenological results} 
\label{sec:Pheno}

As highlighted in Sec.~\ref{sec:intro}, a major advantage of the
\NLOPS simulation over the NLO predictions is that they enable a fully
exclusive simulation of final-state radiation while retaining the NLO
accuracy of the hard scattering process.
In this section we consider full particle level predictions obtained
with our \NLOPS generator interfaced to \PYTHIAn.  We use the same
PDFs, scale settings, and electroweak input parameters specified in
Sec.~\ref{sec:Validation}, but we also include QED radiation and
hadronization effects in the \PYTHIAn{} simulation, which allow us to
provide predictions for the production of hadrons, and to investigate
properties of their distributions.
We note that the inclusion of QED corrections can have important
consequences for the description of charged-lepton based observables
(see the recent discussion in Ref.~\cite{Plestid:2024bva}), and that
the leading corrections are naturally included (and resummed) by the
parton shower in the following.
Specifically, we consider fixed-target collisions on oxygen atoms for
electron neutrinos with energies of $0.1$ and $1~\PeV$, which are
primarily relevant for analyses aiming to measure the flux of cosmic
neutrinos.

\subsection{Particle multiplicities}
\begin{figure*}[ht!]
  \centering
  \begin{subfigure}[h]{.375\textwidth}
    \includegraphics[width=\textwidth]{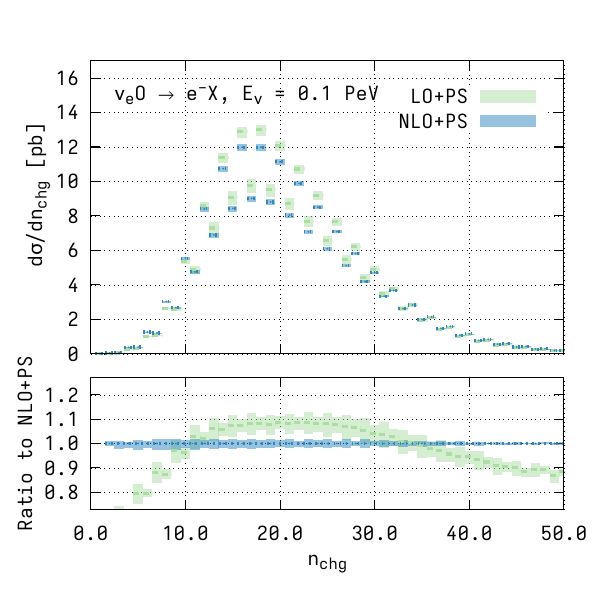}\\[-6ex]
    \subcaption{}
    \label{fig:nchgE5}
  \end{subfigure}
  \hspace{1cm}
  \begin{subfigure}[h]{.375\textwidth}
     \includegraphics[width=\textwidth]{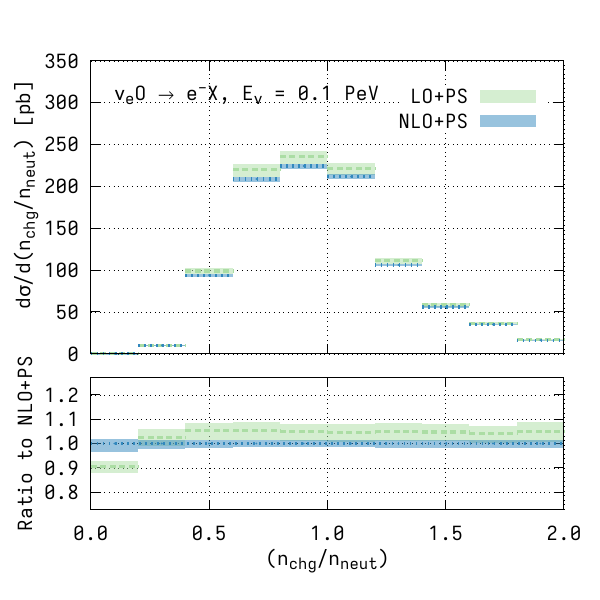}\\[-6ex]
    \subcaption{}
    \label{fig:rchgE5} 
  \end{subfigure}\\[-3ex]
  \begin{subfigure}[h]{.375\textwidth}
    \includegraphics[width=\textwidth]{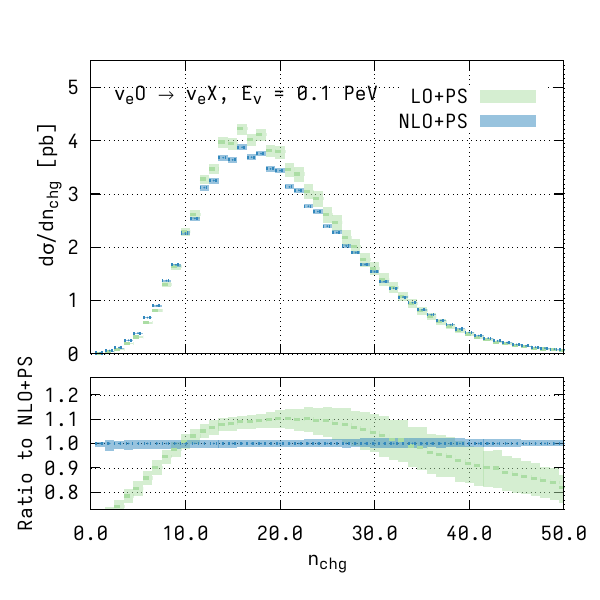}\\[-6ex]
    \subcaption{}
    \label{fig:NCnchgE5}
  \end{subfigure}
  \hspace{1cm}
  \begin{subfigure}[h]{.375\textwidth}
     \includegraphics[width=\textwidth]{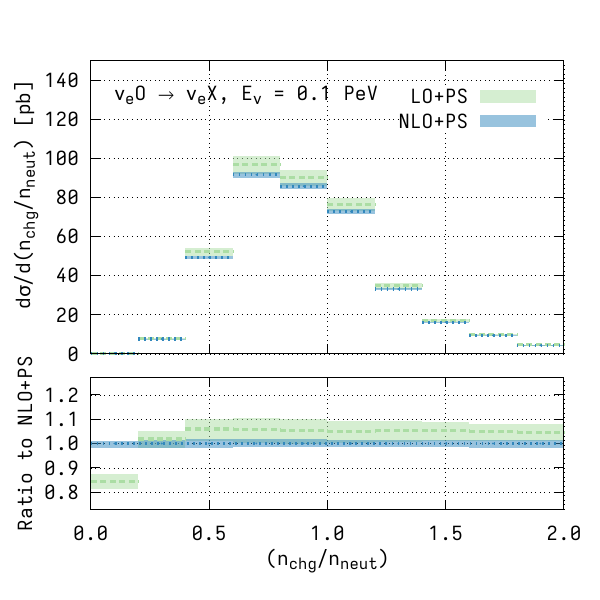}\\[-6ex]
    \subcaption{}
    \label{fig:NCrchE5}    
  \end{subfigure}\\[-3ex]
  \begin{subfigure}[h]{.375\textwidth}
    \includegraphics[width=\textwidth]{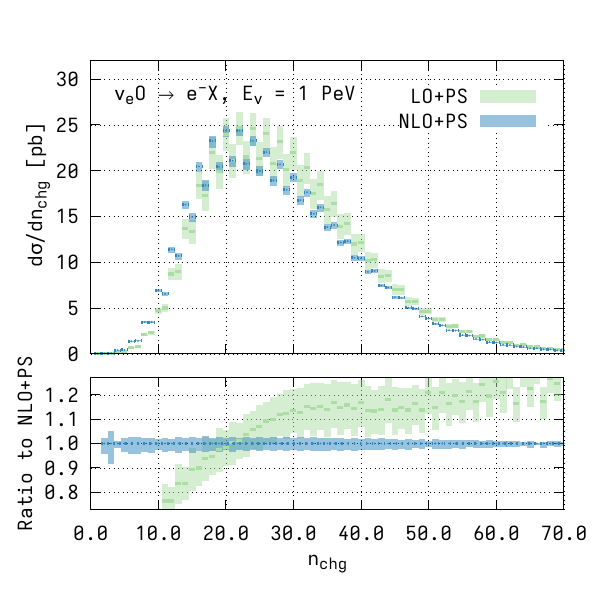}\\[-6ex]
    \subcaption{}
    \label{fig:CCnchgE6}
  \end{subfigure}
  \hspace{1cm}
  \begin{subfigure}[h]{.375\textwidth}
     \includegraphics[width=\textwidth]{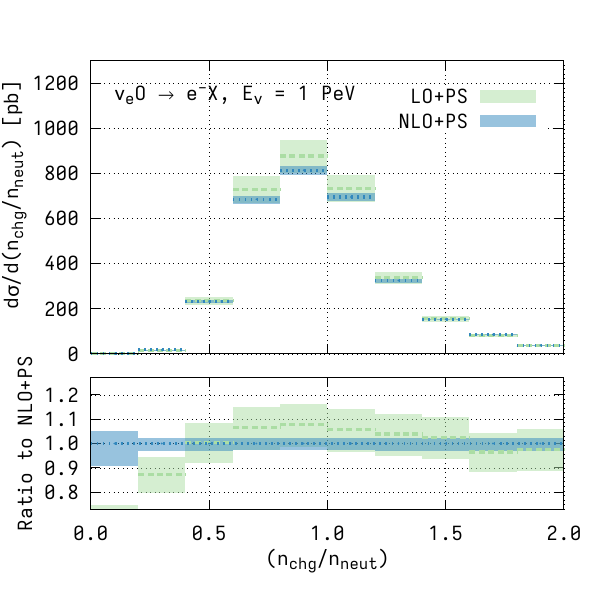}\\[-6ex]
    \subcaption{}
    \label{fig:CCrchE6}    
  \end{subfigure}\\[-1ex] 
  \caption{ Charged particle multiplicity distribution (left) and
    multiplicity ratio between charged and neutral particles (right)
    obtained at \NLOPS (blue) and \LOPS (green) accuracy for neutrino
    induced CC DIS, panels (a),(b), and NC DIS panels (c),(d), on an
    oxygen target with a neutrino energy of $E_{\nu} = 0.1~\PeV$, and
    for CC DIS with $E_{\nu} = 1~\PeV$, panels (e),(f).  The widths of
    the bands indicate scale uncertainties estimated by a 7-point
    variation of $\mu_R$ and $\mu_F$ by a factor of two around the
    central value $Q$. The lower panels show ratios to the respective
    \NLOPS results with $\mu_R=\mu_F=Q$.  }
  \label{fig:multiplicities}   
\end{figure*}
Water-based detector concepts rely on observing the Cherenkov
radiation pattern generated by charged particles in the detector
volume.
An accurate modelling of particle multiplicities in such scattering
events is therefore critical.
Charged particle multiplicities, as well as the ratio of charged to
neutral particle multiplicities are shown in
Fig.~\ref{fig:multiplicities} for $\nu_e$-induced CC and NC DIS at
$E_{\nu} = 0.1~\PeV$, (upper and middle panels), and CC DIS at
$E_{\nu} = 1~\PeV$ (lower panels).
The multiplicity distribution at $E_{\nu} = 0.1~\PeV$ peaks for a number of charged particles, $n_\mathrm{ch}$, of  about 18 
in both the CC and NC cases. At $E_{\nu} = 1~\PeV$
the peak is shifted to around $n_\mathrm{ch}=22$.

As a consequence of charge conservation, an odd (even) number of
charged particles is generated in CC neutrino scattering off protons
(neutrons).  Furthermore, because of the different flavour composition
and associated PDFs of these two types of target particles, the
absolute scattering rate is different for CC on a proton and on a
neutron.  The combination of these effects leads to the observed
``oscillatory'' behaviour for the $n_\mathrm{ch}$ distributions.  This
feature is slightly less pronounced at higher neutrino energies, as
the contribution from PDFs at smaller values of $x$, where the isospin
asymmetric contribution of valence quarks is less important, becomes
more relevant.
We note that the ratio $n_\mathrm{ch}/n_\mathrm{neut}$ peaks at smaller values for the NC process.
Generally, we observe a reduction of scale uncertainty when including
NLO corrections and considerable shape changes induced by NLO effects
which are outside the LO scale uncertainty band, both for the charged
particle multiplicities, as well as the ratios.
When considering higher neutrino energies we notice that the charged
particle multiplicity increases, as expected, and that the NLO
corrections are becoming yet more pronounced and the theoretical
uncertainty stemming from scale variation increases.

It is interesting to note that the centre-of-mass energies considered
here are comparable to those of the HERA collider. Our \NLOPS{}
implementation opens up the opportunity for the re-tuning of event
generators such as \PYTHIAn, which could be relevant given the large
impact of \NLOPS{} corrections on particle multiplicities.

\subsection{Energy-based distributions}
\begin{figure*}[ht!]
  \centering
  \begin{subfigure}[h]{.375\textwidth}
    \includegraphics[width=\textwidth]{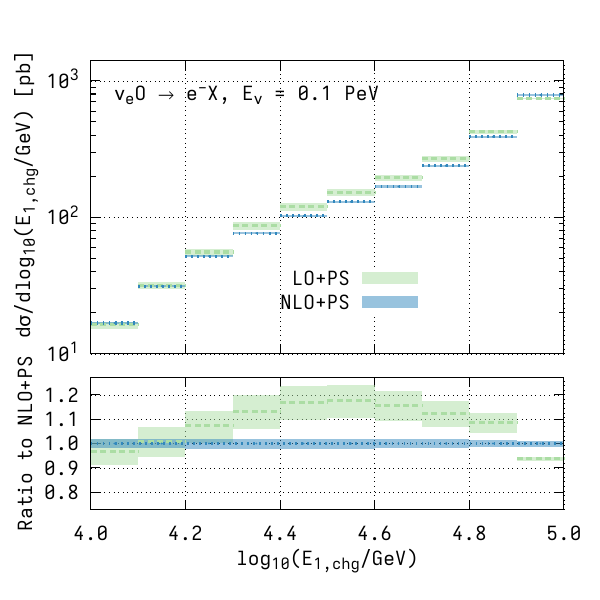}\\[-6ex]
    \subcaption{}
    \label{fig:E1chg5}
  \end{subfigure}
  \hspace{1cm}
  \begin{subfigure}[h]{.375\textwidth}
     \includegraphics[width=\textwidth]{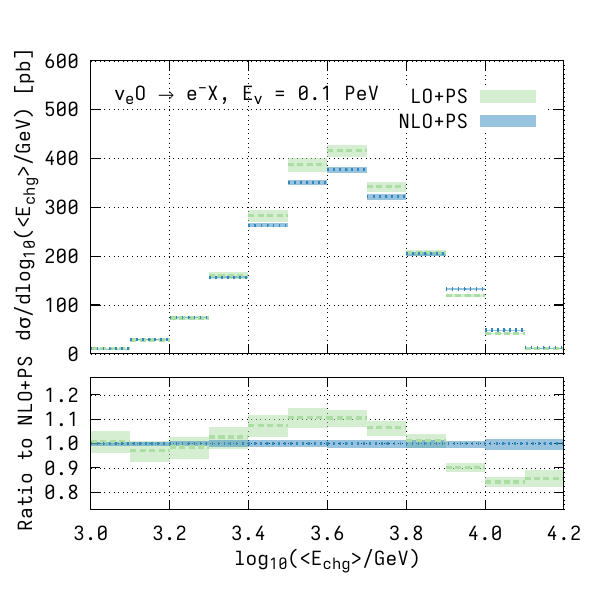}\\[-6ex]
    \subcaption{}
    \label{fig:Emean5}    
  \end{subfigure}
  \\[-3ex]
  \begin{subfigure}[h]{.375\textwidth}
    \includegraphics[width=\textwidth]{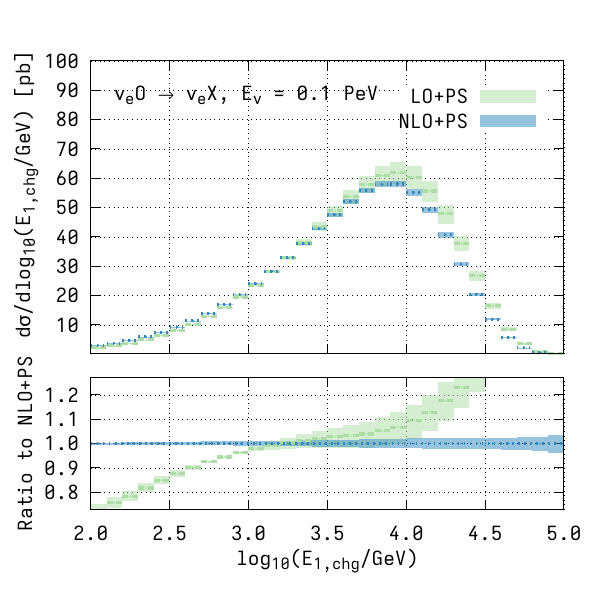}\\[-6ex]
    \subcaption{}
   \label{fig:NCE1chg5}
  \end{subfigure}
  \hspace{1cm}
  \begin{subfigure}[h]{.375\textwidth}
     \includegraphics[width=\textwidth]{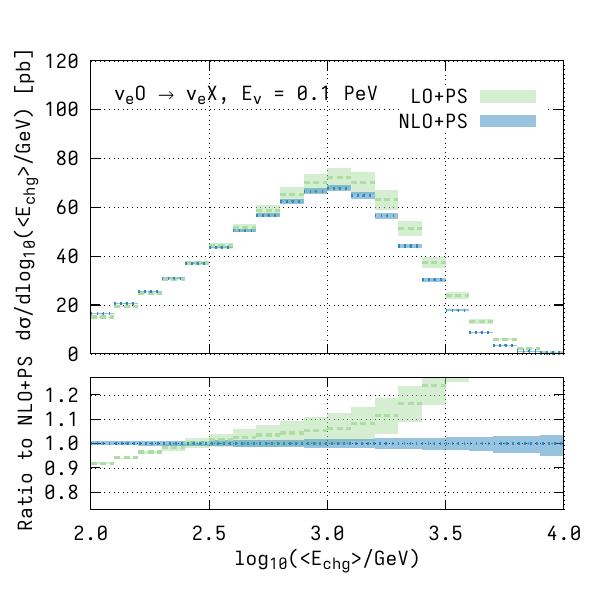}\\[-6ex]
    \subcaption{}
    \label{fig:NCEmean5}    
  \end{subfigure}\\[-3ex]
  \begin{subfigure}[h]{.375\textwidth}
    \includegraphics[width=\textwidth]{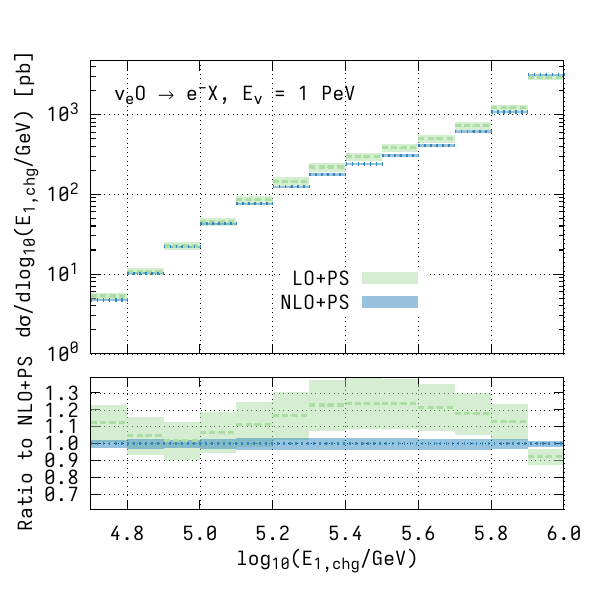}\\[-6ex]
    \subcaption{}
    \label{fig:E1chg6}
  \end{subfigure}
  \hspace{1cm}
  \begin{subfigure}[h]{.375\textwidth}
     \includegraphics[width=\textwidth]{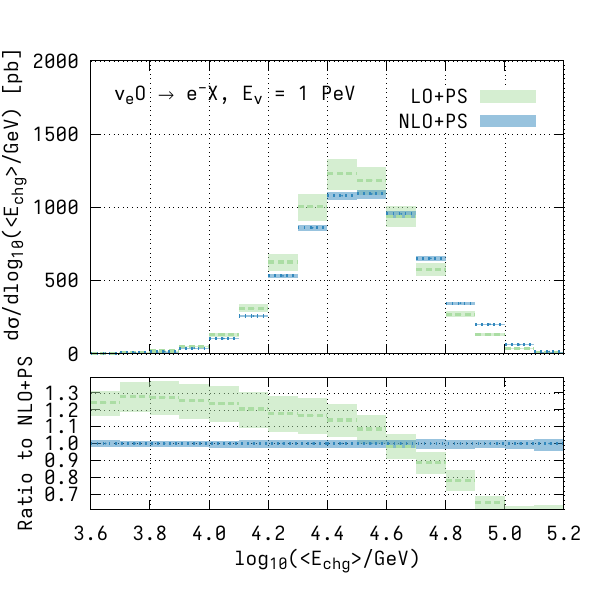}\\[-6ex]
    \subcaption{}
    \label{fig:Emean6}    
  \end{subfigure}
  \label{fig:rchg}\\[-1ex] 
  \caption{Similar to Fig.~\ref{fig:multiplicities}, but for the
    energy of the leading charged particle, $E_{1,\rm chg}$, (left)
    and the mean charged particle energy $\langle E_{\rm chg}\rangle$
    (right).  }
  \label{fig:Echg}
\end{figure*}
In Fig.~\ref{fig:Echg} we compare the predictions for the energy of
the hardest charged particle, $E_{1,\rm chg}$, and the mean charged
particle energy, $\langle E_{\rm chg}\rangle$, as predicted at \LOPS
and \NLOPS accuracy. We notice that these energy distributions are
genuinely different for the CC and NC cases.
This is due to the fact that in the CC case the outgoing lepton
contributes to both distributions, while this is not the case for NC.
For this reason, NLO corrections turn out to be moderate in the CC
case, which is dominated by the lepton kinematics, but considerable
for NC.
We note that, generally, for the determination of $E_{1,\rm chg}$ and
$\langle E_{\rm chg}\rangle$ all charged particles (i.e.\ hadrons and
leptons) are taken into account. If, however, the outgoing charged
lepton is not included in the definition of $E_{1,\rm chg}$ or
$\langle E_{\rm chg}\rangle$ in the CC case, it is observed that the
resultant distributions (and the behaviour of the NLO corrections) are
similar to those of the NC case.
Like for the case of the particle multiplicity, the LO scale
uncertainty band significantly underestimates the size of
higher-order effects as it does not overlap with the NLO band in the
majority of the phase space.
When going to higher energies (plots (e) and (f)), the peaks of the
distributions move accordingly and we find that, as for particle
multiplicities, NLO corrections become more pronounced.

\subsection{Charm production}
\begin{figure*}[th!]
  \centering
  \begin{subfigure}[h]{.40\textwidth}
    \includegraphics[width=\textwidth]{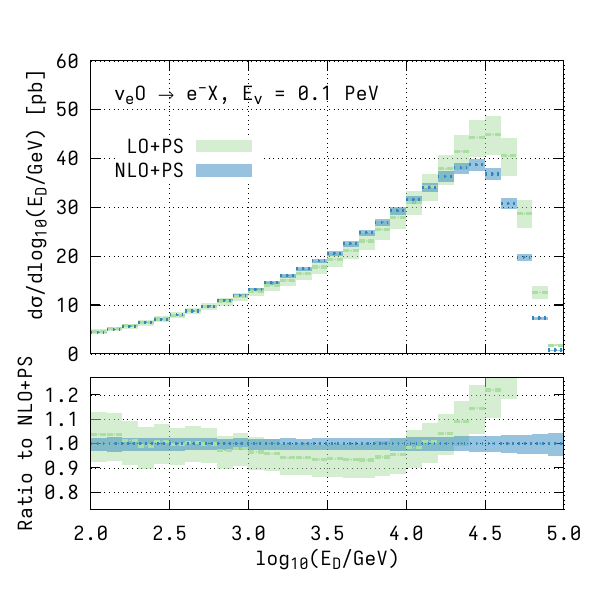}
    \subcaption{}
    \label{fig:ccnchgE5}
  \end{subfigure}
  \hspace{1cm}
  \begin{subfigure}[h]{.40\textwidth}
     \includegraphics[width=\textwidth]{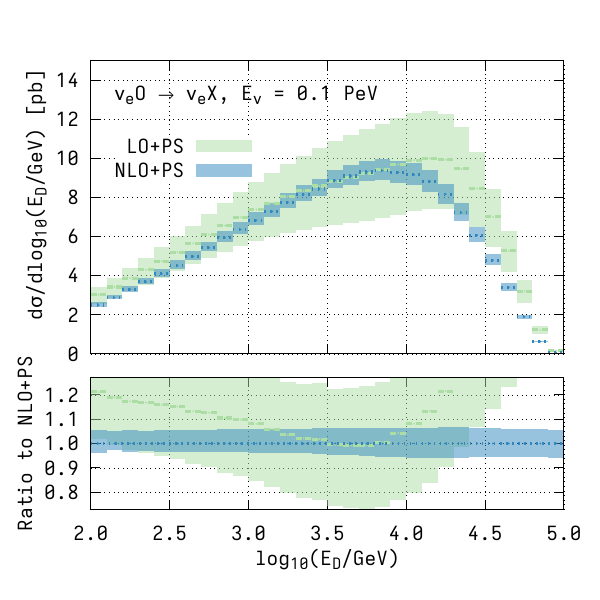}
    \subcaption{}
    \label{fig:ccnchnneutE5}    
  \end{subfigure}
  \begin{subfigure}[h]{.40\textwidth}
    \includegraphics[width=\textwidth]{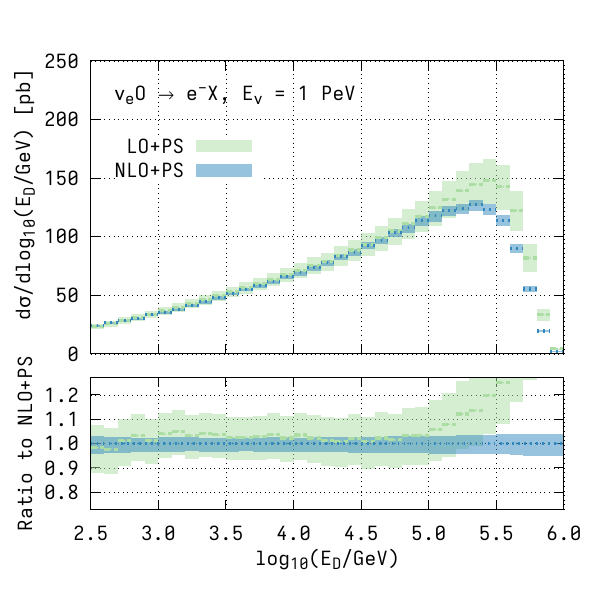}
    \subcaption{}
    \label{fig:ccnchgE6}
  \end{subfigure}
  \hspace{1cm}
  \begin{subfigure}[h]{.40\textwidth}
     \includegraphics[width=\textwidth]{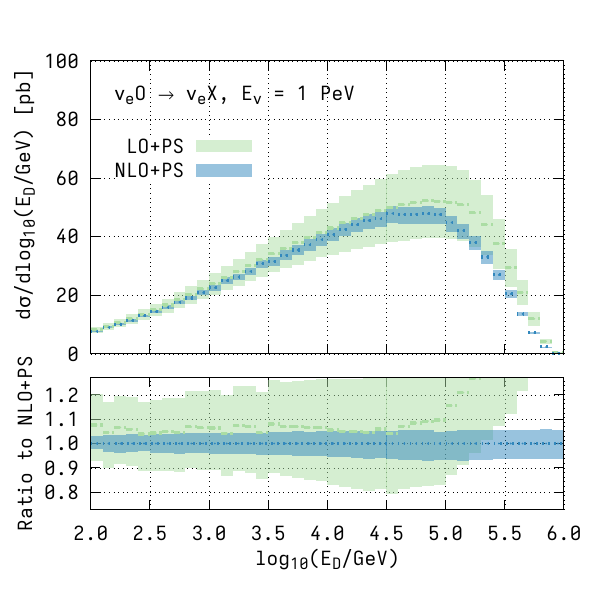}
    \subcaption{}
    \label{fig:ccnchnneutE6}    
  \end{subfigure}
  \caption{ $D$-meson energy distributions at \NLOPS (blue) and \LOPS
    (green) accuracy for neutrino induced CC (left) and NC (right) DIS
    with a neutrino energy of $E_{\nu} = 0.1~\PeV$, panels (a),(b),
    and $E_{\nu} = 1~\PeV$, panels (c),(d).
    The widths of the bands indicate scale uncertainties estimated by
    a 7-point variation of $\mu_R$ and $\mu_F$ by a factor of two
    around the central value $Q$. The lower panels show ratios to the
    respective \NLOPS results with $\mu_R=\mu_F=Q$.  }
  \label{fig:charm}   
\end{figure*}
It is also interesting to investigate the effect of QCD corrections on
$D$-meson distributions.
This is relevant as, through semi-leptonic decays, $D$-mesons provide
a source of energetic muons which can mimic a starting track
signature similar to that arising from muon-neutrino induced CC. 
As discussed in Sec.~\ref{app:compothers}, despite being based on a
purely massless calculation, once interfaced to a parton shower, our
event generator is well suited to describe DIS processes involving
heavy quarks if their mass is much smaller than $Q$, as considered in
this section.
In fact, at the considered neutrino energies, the typical $Q^2$ value
which dominates the cross-section is far in excess of the charm quark
mass (i.e. $|Q^2| \gg m_c^2$), as shown in Fig.~\ref{fig:CCQ2}. In such
a kinematic regime a massless approach to describing the scattering
process is the appropriate one, and ensures a
resummation of the logarithmically enhanced terms in both the initial
and final-state.

We consider here the production of stable $D$-mesons at \LOPS and
\NLOPS accuracy, where the $D$-mesons are produced using the
hadronization feature of \PYTHIAn.
In Fig.~\ref{fig:charm} we present the distribution of the $D$-meson
energy, $E_D$, in the CC and NC cases, respectively.
We find that in the CC case NLO corrections are moderate for low
energies, but become large for high values of $E_D$, where the cross
section peaks.

The CC case is dominated by scattering off $d$- and $s$-quark
distributions, while NC involves primarily a $c$-PDF, which is generated
perturbatively and has a large factorization scale dependence.
For this reason, for NC DIS the scale uncertainties are larger than in the
CC case. These are substantially reduced at NLO.
In each case, the NLO corrections are essential for a reasonable
description of the shape of the energy distribution.

%% file: sec5-Conclusions.tex
\section{Conclusions} \label{sec:Conclusions}

This work presents a number of extensions to the simulation of
neutral- and charged-current deep inelastic scattering
(DIS)~\cite{Banfi:2023mhz} in the \RES{}.
First, the code has been extended to accommodate an incoming neutrino
beam. Second, the incoming lepton is no longer required to be
monochromatic, as in standard high-energy DIS experiments.
Instead, any incoming lepton flux can be included.
Moreover, an option is provided to straightforwardly account for the
kinematics of fixed-target experiments.
Furthermore, more flexible options for the nuclear targets are now
supported.

With the new implementation we have provided sample results for
fiducial cross-sections, standard DIS variables, as well as neutral
and charged particle distributions for various neutrino-induced DIS
processes.
In our sample numerical analyses we put a particular focus on the
kinematic regime relevant for the investigation of cosmic neutrinos
with the IceCube detector.
We note, however, that our program is not restricted to this
application, but can be employed for the simulation of any
neutrino-induced DIS process.
In general, we find that an \NLOPS simulation is necessary to achieve
theory uncertainties below approximately 10\%.

The code, along with the new features discussed in this article, is
publicly available via the \RES repository.
The reliance on the \RES{} framework, which is well-suited for
describing complex reactions involving multiple competing and
interfering sub-processes, will enable us to further improve the
description of hadron-collider processes such as vector boson
scattering and vector boson fusion, going beyond what is already
available in \POWHEGBOXVTWO{}.
These reactions can be described as (generalized) two-fold DIS
processes, and are highly relevant for the phenomenology of the Large
Hadron Collider.
Additionally, our approach paves the way for the simulation of
electroweak corrections in DIS consistently accounting for photon
radiation in the hadronic and leptonic sectors.

%% file: appendix.tex

\section[\phantom{Appendix} DIS process selection]{DIS process selection}
\label{sec:processes}

In this appendix we summarise inputs that can be used to select the
process and settings in the \texttt{powheg.input} file, which are
specific to the DIS case.

\paragraph{\textbf{Lepton beam}.}
The flavour of the incoming lepton must be specified using
\verb|ih1 int|, 
where  the integer number  \texttt{int} is the identifier of the desired lepton in the
{\em Particle Data Group} numbering convention~\cite{ParticleDataGroup:2022pth}.

The energy of the lepton beam must be specified using \texttt{ebeam1
double} with a double-precision number \texttt{double}.  By default
the code assumes a fixed lepton energy. To use a variable flux add the
option\\
\verb|fixed_lepton_beam 0|, 
(see Sec.~\ref{sec:neutrino-flux} for more details).
A variable flux should be provided in terms of a boost-invariant
energy fraction, of the lepton beam with repect to the maximum energy
available.

\paragraph{\textbf{Hadron beam/target}.}
The selection of the nucleon type in the \texttt{powheg.input} file
must be chosen by setting the value of \texttt{int} in \verb|ih2 int|.
We currently support protons and neutrons.  To that end the following
options are available:
\begin{enumerate}
	\item[] \verb|ih2 1  #proton target, input proton PDF| 
	\item[] \verb|ih2 2  #neutron target, input proton PDF|
	\item[] \verb|ih2 22 #neutron target, input neutron PDF|
\end{enumerate}
Depending on the selection for \texttt{ih2}, the PDF specified via the
entry \texttt{lhans2} according to the numbering scheme of the LHAPDF
repository~\cite{Buckley:2014ana} is interpreted either as a proton or
a neutron PDF.\footnote{\texttt{lhans1} must be set to the same value
as \texttt{lhans2}, even if not used, in case the PDF implementation
of the running of the QCD coupling constant (\texttt{alphas\_from\_pdf
1}) is to be used.}

The energy of the hadron is selected via the mandatory entry
\texttt{ebeam2 double}.  By default, the code assumes that the hadron
beam is massless, with a longitudinal momentum equal to its energy.
For fixed-target collisions, one has to add the option
\begin{verbatim}
fixed_target 1
\end{verbatim}
In this case the value of the entry for \texttt{ebeam2} is interpreted
as the mass of the nucleon (i.e. proton or neutron).

\paragraph{\textbf{Hard process selection}.}
Both CC and NC processes can be simulated within our framework.  To
select the desired channel (for a given type of lepton beam, as
specified by the value of \texttt{ih1}), one can use the following
option
\begin{verbatim}
channel_type int
\end{verbatim}
with \texttt{int}=3 for CC, and \texttt{int}=4 for NC.
In case of charged-lepton induced NC DIS, the boson exchanged in the
$t$-channel has to be specified using \verb|vtype| with
\begin{enumerate}
\item \verb|vtype 1 # photon exchange only|
\item \verb|vtype 2 # Z echange only|
\item \verb|vtype 3 # photon+Z exchange|
\end{enumerate}

\paragraph{\textbf{Generation cuts}.}
The user must specify cuts on the DIS invariants $Q^2$, $\xB$ and
$\ydis=Q^2/(\xB S)$, with $S=(P_1+P_2)^2$.  The values
of \texttt{Qmin} and \texttt{Qmax} are supposed to be provided in
units of GeV.  For example, to probe all the available phase space,
one should set
\begin{verbatim}
Qmin 1d0
Qmax 1d8
xmin 0d0
xmax 1d0
ymin 0d0
ymax 1d0
\end{verbatim}
where \texttt{Qmax} has been set to a value much larger than the
center-of-mass energy.
We stress that \texttt{Qmin}=1~GeV is the lowest value accepted by the
code,
since the validity of a perturbative QCD approach to describe the
cross section is no longer guaranteed for small $Q^2$.

We note that it is possible to fix up to 2 of these variables by
setting the minimum and maximum values equal to each other.
In any case, the code will never generate events outside the physically
allowed bounds.

\paragraph{\textbf{Final-state particles masses}.}
Notice that all particles entering the hard process are treated as
massless in our NLO calculation.
As in most \POWHEGBOX{} implementations, a small reshuffling of the
momenta can be applied when generating events, so as to give a finite
mass to all the final-state massive particles.
The mass of the charged leptons and of the heavy quarks can be
specified, e.g.\ using
\begin{verbatim}
electron_mass 0.51099891d-3 
muon_mass 0.1056583668d0 
tauon_mass 1.77684d0
charm_mass 1.5d0
bottom_mass 4.5d0
\end{verbatim}
The numbers chosen in this example correspond to the default values.
More comments on how mass effects are approximately included in our
generator are given in~\ref{app:compothers}.